\documentclass[aps,prl,twocolumn,superscriptaddress]{revtex4-2}
\usepackage[utf8]{inputenc}
\usepackage[american,british]{babel}
\usepackage[T1]{fontenc}
\usepackage[pdftex]{graphicx}  
\usepackage{graphicx, xcolor}
\usepackage{dcolumn}
\usepackage{bm}
\usepackage{amsmath,amsthm,amssymb}
\usepackage{color}
\usepackage{verbatim}

\usepackage{braket}

\definecolor{darkGreen}{RGB}{0,110,0}
\definecolor{darkBlue}{RGB}{0,0,130}
\usepackage[colorlinks,citecolor=darkGreen,linkcolor=darkBlue,%
urlcolor=blue,hyperindex]{hyperref}

\usepackage{ulem} 
\renewcommand{\vec}{\boldsymbol}

\usepackage{bm}

\begin{document}

\title{Highly resolved spectral functions of two-dimensional systems with neural quantum states}

\author{Tiago Mendes-Santos}
\affiliation{Theoretical Physics III, Center for Electronic Correlations and Magnetism,
Institute of Physics, University of Augsburg, 86135 Augsburg, Germany}
\author{Markus Schmitt}
\affiliation{Forschungszentrum J\"ulich GmbH, Peter Gr\"unberg Institute, Quantum Control (PGI-8), 52425 J\"ulich, Germany}
\author{Markus Heyl}
\affiliation{Theoretical Physics III, Center for Electronic Correlations and Magnetism,
Institute of Physics, University of Augsburg, 86135 Augsburg, Germany}

\begin{abstract}
Spectral functions are central to link experimental probes to theoretical models in condensed matter physics.
However, performing exact numerical calculations for interacting quantum matter has remained a key challenge especially beyond one spatial dimension. 
In this work, we develop a versatile approach using neural quantum states to obtain spectral properties based on
simulations of the dynamics of excitations initially localized in real or momentum space.
We apply this approach to compute the dynamical structure factor in the vicinity of quantum critical points (QCPs) of different two-dimensional quantum Ising models, including one that describes the complex density wave orders of Rydberg atom arrays.
When combined with deep network architectures 
we find that our method reliably describes dynamical structure factors of arrays with up to $24\times24$ spins, including the diverging time scales at critical points. 
Our approach is broadly applicable to interacting quantum lattice models in two dimensions and consequently opens up a route to compute spectral properties of correlated quantum matter in yet inaccessible regimes.
\end{abstract}

\maketitle

\textit{Introduction.} 
Spectral functions are key tools to characterize and probe quantum many-body phases and their transitions.
In addition, they serve as a common framework to connect theoretical descriptions with experimental probes such as photoemission or inelastic neutron scattering. 
In this context, a regime of  particular interest is two-dimensional (2D) interacting quantum matter, where experimental probes can indicate the occurrence of prominent properties, such as exotic fractionalized quasiparticles in
candidate materials realizing 2D spin-liquid phases \cite{Balents2010} or universal features associated to quantum critical points \cite{Sachdev2011}.

\begin{figure}[]
{\centering\resizebox*{8.5cm}{!}{\includegraphics*{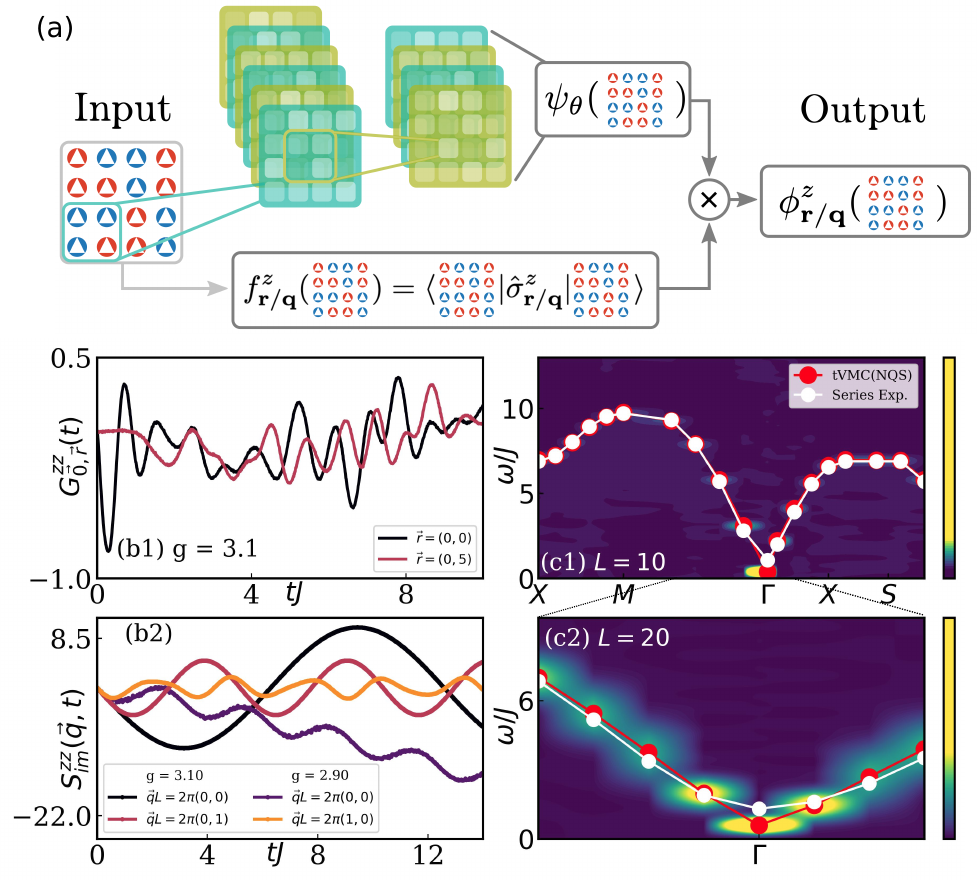}}}
\caption{
Panel (a) shows a schematic picture of the NQS architecture used in this work. Besides the convolutional neural network $\psi_{\mathbf\theta}(\mathbf s)$ that forms the variational part, the initial excitations in real or momentum space ($\mathbf r$ or $\mathbf q$) are captured by directly incorporating the corresponding operator matrix elements $f_{\mathbf r/\mathbf q}(\mathbf s)=\braket{\mathbf s|\hat\sigma^z_{\mathbf r/\mathbf q}|\mathbf s}$ in the ansatz. The logarithmic wave function coefficients of the state $\ket{\phi^{z}_{\vec{r}/\vec{q}}}$ then take the form $\log\braket{\mathbf s|\phi^{z}_{\vec{r}/\vec{q}}}=\log\psi_{\mathbf\theta}(\mathbf s)+\log f_{\mathbf r/\mathbf q}(\mathbf s)$ as indicated by the computational graph.
Panels (b1,c1)  show some exemplary results of real-space dynamical correlators and the corresponding DSF 
 close to the quantum critical point of the two-dimensional quantum Ising model.
Analogously, panels (b2,c2) show results obtained directly in momentum space for larger system sizes. We consider the path $X = (\pi,0) \to M = (\pi,\pi) \to \Gamma = (0,0) \to X \to S=(\pi/2,\pi/2)$ of the Brilloin Zone.}
\label{fig1} 
\end{figure}

At the theoretical level, accessing and describing spectral functions is, thus, of great interest in strongly interacting solid-state materials. But addressing dynamical properties of correlated matter in a controlled manner poses, at the same time, substantial challenges. Quantum Monte Carlo is poised by a sign problem \cite{Troyer2005} and the applicability of dynamical mean field theory \cite{RevModPhys.86.779} is limited in low dimensions. Tensor network approaches, which render the treatment of weakly entangled states feasible, can be used to obtain numerically exact results for one-dimensional systems \cite{Hallberg1995,Kuhner1999,Jeckelmann2002,Benthien2004,White2004,Paeckel2019}. While extensions to higher dimensions exist \cite{Zaletel2015,Gohlke2017,Verresen2018,Verresen2019,VanDamme2022}, the growth of entanglement in time together with the two-dimensional lattice structure that increases the complexity of tensor contractions remains as a challenge for tensor network methods. 
In addition,  variational methods for
capturing excitations based on Gutzwiller-projected mean-field states are restricted to specific cases due to their built-in bias \cite{Becca2018,Becca2019}.
Finally, programmable quantum simulation could emerge as a new route \cite{Knap2013,Baez2020,Kim2023}, but it is still in its infancy at this point.

Recently, the idea to combine the variational Monte Carlo (VMC) framework with neural quantum states (NQSs) \cite{Carleo2017} has been shown to be very fruitful for investigations of correlated matter, including the simulation of
ground states of frustrated Hamiltonians \cite{Imada2021,nomura2021,Astrakhantsev2021,roth2022,reh2023,chen2023} and 
the dynamics of two-dimensional systems \cite{Schmitt2020,Reh2021,Fabiani2021,Schmitt2022_qkzm,donatella2022,mendessantos2023}.
For spectral functions, first attempts proposed NQS-based algorithms built directly in the frequency domain \cite{Neupert2018,Feiguin2019,Feiguin2021} or a method simulating the response to an initial time-dependent perturbation of the system \cite{Mentink2019}. However, it remains desirable to enhance the resolution and reachable system sizes over what has been achieved so far in order to address open physical questions.

In this work, we introduce an alternative versatile scheme for the simulation of spectral functions based on the direct encoding of local excitations in the neural network architecture -- the specNQS, see Fig.~\ref{fig1}(a).
The spectral information is then extracted from dynamical correlation functions that are obtained by real-time evolution.
When combined with convolutional neural networks, we demonstrate that our scheme allows us to access dynamical properties beyond what has been feasible with other state-of-the-art approaches.
As a benchmark, we simulate the dynamical structure factor (DSF) of the  2D quantum  Ising model (QIM), and we showcase that the specNQS reliably describes spectral features associated with a diverging correlation length for system sizes up to  $24 \times 24$ sites.
Furthermore, we contribute to the characterization of quantum phase transitions in experimentally realized long-range interacting Rydberg atom arrays \cite{Ebadi2021,Scholl2021} by revealing spectral properties close to phase boundaries, the nature of which is under ongoing debate \cite{Felser2021,Kalinowski2022}.

\textit{Method.}
In the following, we will be interested in computing the DSF
\begin{align}
    S^{zz}(\vec q,\omega)
    &=
    \frac{1}{N_s} \int_{-\infty}^\infty dte^{i\omega t}
    \big\langle\hat \sigma^z_{-\vec q}(t)\hat \sigma^z_{\vec q}\big\rangle.
    \label{eq:DSF}
\end{align}
Here, $\langle\hat A(t)\hat B\rangle=\braket{\psi_0|e^{i\hat H t} \hat A e^{-i\hat H t}  \hat B|\psi_0}$ denotes the  dynamical correlation function in the ground state $\ket{\psi_0}$ of a given Hamiltonian $\hat H$, $\hat H\ket{\psi_0}=E_0\ket{\psi_0}$ with $E_0$ being the ground-state energy. 
$\hat \sigma^z_{\vec q}=\sum_{\vec r}e^{-i\vec q\cdot\vec r}\hat \sigma^z_{\vec r}$ is the spin operator in momentum space, where $\hat \sigma^z_{\mathbf r}$ denotes the Pauli-$z$ operator at lattice site $\vec r$ and $N_s$ is the number of lattice sites. 

The central idea of our approach is to obtain the DSF from a variational representation of suitably time-evolved wave functions. NQSs constitute a versatile family of variational wave functions relying on the proven representational power of artificial neural networks (ANNs).
In particular, any function can be accurately approximated by an ANN in the limit of large network sizes \cite{Cybenko1989, Hornik1991, Kim2003, LeRoux2008}. 
This means that the accuracy of the proposed approach  can be asserted self-consistently by checking the convergence with increasing network size.

As a first step to access the DSF, we compute  the NQS representation of the ground state 
\begin{equation}
 \ket{\psi_0} = \sum_{\vec s}  \psi_{\vec{\theta}}^{(0)}(\vec s) \ket{\vec s}, 
 \label{excitation}
\end{equation}
where $\vec s = (s_1, s_2, ..., s_{N_s})$ labels the Pauli-Z basis of spin configurations.
The variational ansatz $\psi_{\vec{\theta}}^{(0)}(\vec s)$ is parameterized by $\vec{\theta} = (\theta_1,\theta_2, ..., \theta_M)$ and it takes the form of an ANN. 
The ground state is then obtained by optimizing $\vec{\theta}$ to minimize the energy expectation value $\mathcal{E}(\vec{\theta}) =\braket{\psi(\vec\theta)|\hat H|\psi(\vec\theta)}/\braket{\psi(\vec\theta)|\psi(\vec\theta)}$.
For the results presented throughout this manuscript, we employed the Stochastic Reconfiguration algorithm to find the ground states \cite{Sorella1998,Sorella2001}.

Our approach to access the dynamics relies on computing the time-evolved wave functions following an excitation,
\begin{align}
    \ket{\phi_{\vec r / \vec q}^{\alpha}(t)}
    =e^{-i\hat H t}\hat \sigma_{\vec r / \vec q}^{\alpha}\ket{\psi_0}
    \equiv e^{-i\hat H t}\ket{\phi_{\vec r / \vec q}^\alpha(0)}
    \ .
\end{align}
Here, $\hat \sigma_{\vec r / \vec q}^{\alpha}$ is an operator in either position ($\vec r$) or  momentum space ($\vec q$).
In our numerical approach we rely in either case on an exact representation of $\ket{\phi_{\vec r / \vec q}^\alpha(0)}$, which will be discussed in the following paragraphs. Let us first discuss the time-evolution algorithm, assuming that the representation of the initial state is given. We employ a time-dependent variational principle (TDVP) that is based on the minimization of the  Fubini-Study distance between the variational time-evolved state $\ket{\psi_{\vec{\theta}(t + \delta t)}}$ and the exact one $e^{-i\hat H \delta t}\ket{\psi(t)}$, where $\delta t$ is an infinitesimal time interval. It yields an ordinary non-linear differential equation prescribing the optimal evolution of the variational parameters \cite{Carleo2017, Schmitt2020},
\begin{align}
    S_{k,k'} \dot{\theta}_{k'}=-iF_k\ ,
    \label{eq:tdvp}
\end{align}
where $\dot{\theta}_{k'}$ is the time derivative, $S_{k,k'}=\braket{\partial_{\theta_k}\psi_{\vec\theta}|\partial_{\theta_{k'}}\psi_{\vec\theta}}-\braket{\partial_{\theta_k}\psi_{\vec\theta}|\psi_{\vec\theta}}\braket{\psi_{\vec\theta}|\partial_{\theta_{k'}}\psi_{\vec\theta}}$ is the quantum metric tensor and $F_{k}=\braket{\partial_{\theta_k}\psi_{\vec\theta}|\hat H|\psi_{\vec\theta}}-\braket{\partial_{\theta_k}\psi_{\vec\theta}|\psi_{\vec\theta}}\braket{\psi_{\vec\theta}|\hat H|\psi_{\vec\theta}}$. Both $S_{k,k'}$ and $F_k$ can be estimated efficiently via Monte-Carlo sampling of the Born distribution $|\psi_{\vec\theta}(s)|^2$.

Upon integration, the TDVP in Eq.~\eqref{eq:tdvp} in general yields the time-evolved state up to a global phase and normalization \cite{Cirac2020}.
The phase is irrelevant for equal-time correlation functions. For computing the dynamical correlation functions, however, it becomes important, because we are interested in evaluating the overlap of two time-evolved states.
To keep track of relative changes in phase  between such time-evolved states we consider the (logarithmic) prefactor $\theta_0$ as an additional variational parameter: $\ket{\psi_{\theta_0,\vec\theta}}=e^{\theta_0}\ket{\psi_{\vec\theta}}$. 
By using the TDVP, one can establish the following  equation of motion for $\theta_0$ \cite{carleo_phd}
\begin{equation}
 \dot{\theta}_0 = -i \braket{\hat H} - \dot{\theta}_{k} \braket{\psi_{\vec\theta}|\partial_{\theta_k}\psi_{\vec\theta}}.
 \label{eq:eta0}
\end{equation}
Equation \eqref{eq:tdvp} for the other parameters $\vec\theta$ remains unchanged. 
Thus, for each time step, we first obtain $\dot{\theta}_{k}$ and then use the result to solve Eq. \eqref{eq:eta0} for the evolution of $\theta_0$.

\textit{Momentum-space scheme.}
First, we discuss how to access the dynamical spin structure factor directly in momentum space,
\begin{equation}
  S^{zz}(\vec{q},t) = e^{iE_0 t} \bra{\psi_0} \hat \sigma_{\vec{q}}^z  e^{-i \hat{H} t/2}   e^{-i \hat{H} t/2} \hat \sigma_{\vec{q}}^z \ket{\psi_0}\ .
  \label{localob}
\end{equation}
To simplify the discussion we focus on the $z$ component, but our approach can be straightforwardly generalized for $\alpha=x,y$ by choosing the computational basis accordingly.
The central idea is that the action of the operator $\hat \sigma_{\vec q}^z$ on the initial state $\ket{\psi_0}$ can be captured explicitly and efficiently by modifying the individual wave function coefficients with corresponding prefactors.
Concretely, the excitation $\sigma_{\vec{q}}^z \ket{\psi_0}$ is encoded on the NQS ansatz 
by adding a configuration-dependent factor $f(\vec{q},\vec s) = \sum_{j} e^{-i \vec{r}_j  \vec{q}} e^{i\pi/2(s_{j} -1)}$ on the initial-time quantum state
\begin{equation}
 \ket{\phi_{\vec{q}}^{z}(0)} = \hat \sigma_{\vec{q}}^z \ket{\psi_0} = \sum_{\vec s} f(\vec{q},\vec s) \psi_{\vec{\theta}}(\vec s) \ket{\vec s}\ .
 \label{excitation}
\end{equation}
We then  obtain $S^{zz}(\vec{q},t)$  by performing a two-sided  time evolution with the TDVP approach, 
followed by  evaluating the overlap of the two time evolved states,
\begin{equation}
  S^{zz}(\vec{q},t) = e^{iE_0t}\braket{\phi^z_{\vec{q}}(-t/2) | \phi^z_{\vec{q}}(t/2)}.
  \label{eqOverlap}
\end{equation}
This is a the central object to calculate the dynamical structure factor by means of our NQS approach. Notice that this implies that for the dynamical correlation function up to time $t$ numerical integration is only required up to time $t/2$.
Here, it is important to account for the global factor associated to each state of the overlap as discussed above, see Eq.~\eqref{eq:eta0}. Refs~\cite{SM,Wu2020} presents further details about the calculation of the overlap.

\textit{Real-space scheme.}
Second, we discuss a strategy to obtain dynamical correlations in real space.
Our scheme is based on a many-body Ramsey protocol \cite{Knap2013}, that is used to simulate the retarded Green's function (GF) 
\begin{equation}
    G^{\alpha\alpha}_{\vec r,\vec r'}(t) = - \frac{i}{2} \bra{\psi_0} [\hat \sigma_{\vec r}^{\alpha}(t), \hat \sigma_{\vec r'}^{\alpha}(0)] \ket{\psi_0},
    \label{eq:GF}
\end{equation}
where $\vec r$, $\vec r'$ are the sites of a lattice,  and $[\hat A,\hat B]$ denotes the commutator. 
In particular, to access the longitudinal $G^{zz}_{\vec r,\vec r'}(t)$ component,  we start the protocol with the following  quantum state  
$\ket{\phi_{\vec r}^z(0)} = e^{i\frac{\pi}{4} \hat \sigma^z_{\vec r}} \ket{\psi_0}$,
where the local pertubation is  represented by the  $\frac{\pi}{4}$-rotation of a spin at site $\vec r$. 
Further, we obtain the time-evolved state $\ket{\phi_{\vec r}^z(t)} = e^{-i\hat Ht} \ket{\phi_{\vec r}^z(0)} $.
Following Ref. \cite{Knap2013}, the result of a local measurement of $\hat \sigma_{\vec r}^z$ at a time $t$ depends on the GF, i.e.,
\begin{align}
  &\bra{\phi_{\vec r'}^z(t)}\hat \sigma_{\vec r}^z \ket{\phi_{\vec r'}^z(t)} =
  \nonumber\\
  &
  \bra{\psi_0}\hat \sigma_{\vec r}^z(t)\ket{\psi_0}
  + G^{zz}_{\vec r,\vec r'}(t) 
  + \bra{\psi_0}\hat \sigma_{\vec r}^z\hat \sigma_{\vec r'}^z(t)\hat \sigma_{\vec r}^z \ket{\psi_0}.
  \label{localob}
\end{align}
The $G^{zz}_{\vec r,\vec r'}(t)$ is obtained by reconstructing the terms of the Eq. \eqref{localob}.
The first term on the right hand side is accessed from the ground state. For the remaining contributions we time-evolve the initial states $e^{i\frac{\pi}{4}\hat \sigma^z_{\vec r}} \ket{\psi_0}$ and $\hat \sigma_{\vec r}^z \ket{\psi_0}$ after incorporating the operator action into the variational ansatz in analogy to Eq.~\eqref{excitation}.

One central difference between the momentum- and the real-space approach is that the latter does not require the calculation of state overlaps. Moreover, they differ in the way translational symmetry can be exploited for the efficiency of simulations. In the real space approach all correlation functions in Eq.~\eqref{localob} depend only on relative positions $\vec r'-\vec r$, which means that all the momentum points of $S^{zz}(\vec{q}, \omega)$ can be obtained from the two time-evolved states $\ket{\phi_{\vec r=0}^z(t)}$ and $e^{-i\hat H t}\hat \sigma_{\vec r=0}^z \ket{\psi_0}$. However, with this approach, translational symmetry cannot be built into the variational ansatz to enhance efficiency. By contrast, the time-evolved states in Eq.~\eqref{eqOverlap} for the momentum-space approach preserve translational symmetry, which can be exploited to introduce beneficial bias through built-in invariance of the NQS. This comes, however, at the cost of individual simulations required for each point in momentum space.

Finally, it is worth mentioning that we compute $S^{zz}(\vec{q},\omega)$ by performing a Fourier transform with a Gaussian envelope to avoid the finite-time effects of our simulation (our simulations are performed up to a time $t_{max}$) \cite{SM}.

\textit{Neural quantum states architectures.}
In conjunction with the momentum-space scheme, we employ convolutional neural networks (CNNs) \cite{Yoav2020,Schmitt2020} as the variational part of the NQS architecture, which allows us to exploit the translational symmetry, that is also conserved after applying the operators in momentum space.
The hyperparameters of the CNNs are the total number of layers $l$, the number of channels in each layer $k$, $\alpha_k$, and the linear size of the square filter, $F$; in the following, we characterize the CNN architecture with the tuples $ \vec{\alpha} = (\alpha_1,...,\alpha_l; F)$. 
Meanwhile, to implement the real-space scheme, we use Restricted Boltzmann Machines (RBMs) \cite{Carleo2017} composed of a single fully-connected hidden layer with $M$ nodes, where $M$ is a hyperparameter; further details about the NQSs are discussed in the SM \cite{SM}.

\begin{figure}[t]
\begin{center}
{\centering\resizebox*{8.5cm}{!}{\includegraphics*{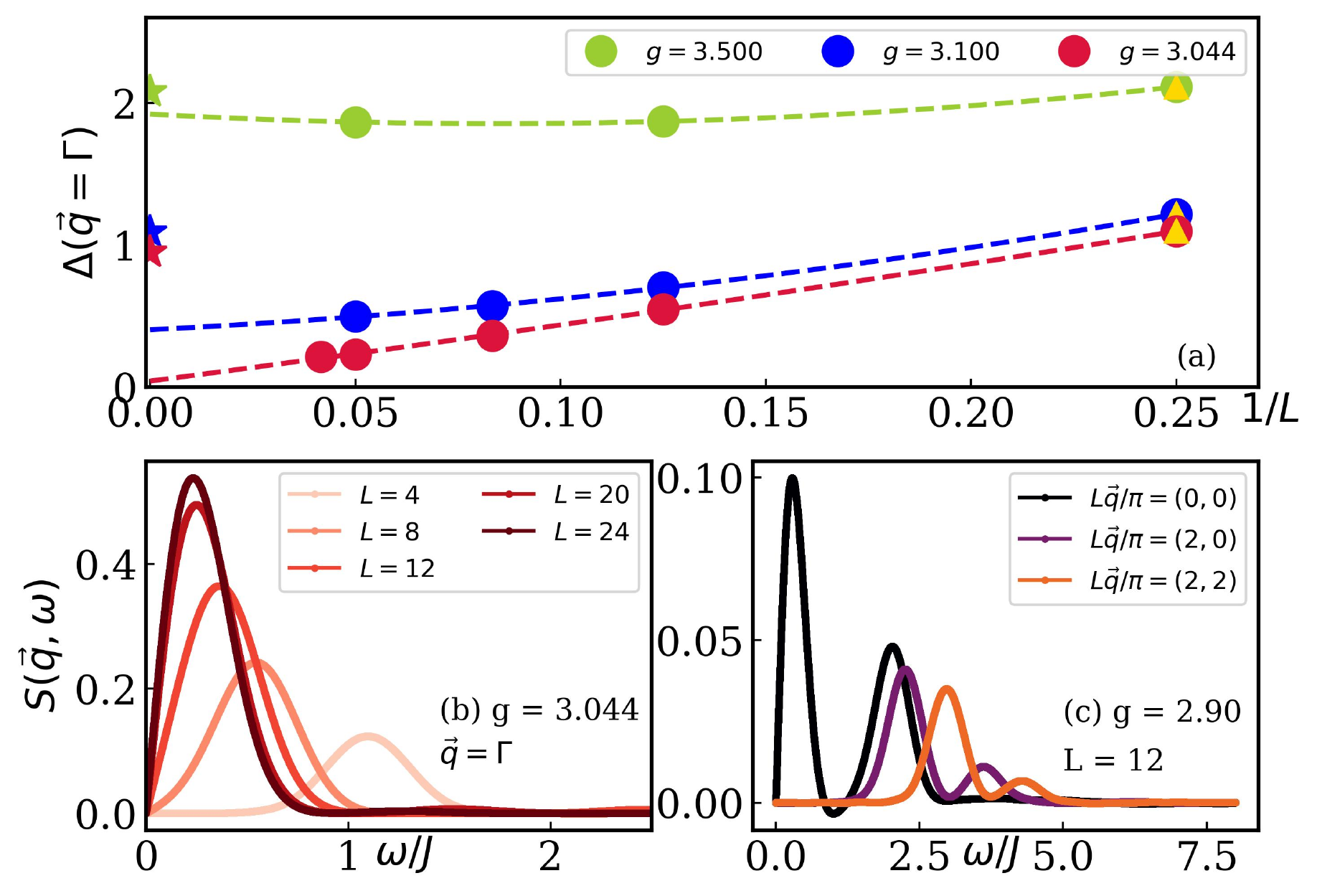}}}
\end{center}
\caption{\textit{2D quantum Ising model, vicinity of the quantum critical point.} 
   Panel (a) shows the finite-size scaling of the gap $\Delta(\vec{q}= \Gamma)$; the star points are series-expansion results up to forth order,
   while the triangle points are exact-diagonalization results for $L =4$.
   In panel (b), we  show the corresponding DSF $S(\vec{q} = \Gamma, \omega)$ at QCP for different values of $L$. 
   In panel (c) we present  $S(\vec{q}, \omega)$ at the ferromagnetic side of the quantum phase transition.
   }
\label{fig3} 
\end{figure}

\begin{figure*}[t]
\begin{center}
{\centering\resizebox*{16.5cm}{!}{\includegraphics*{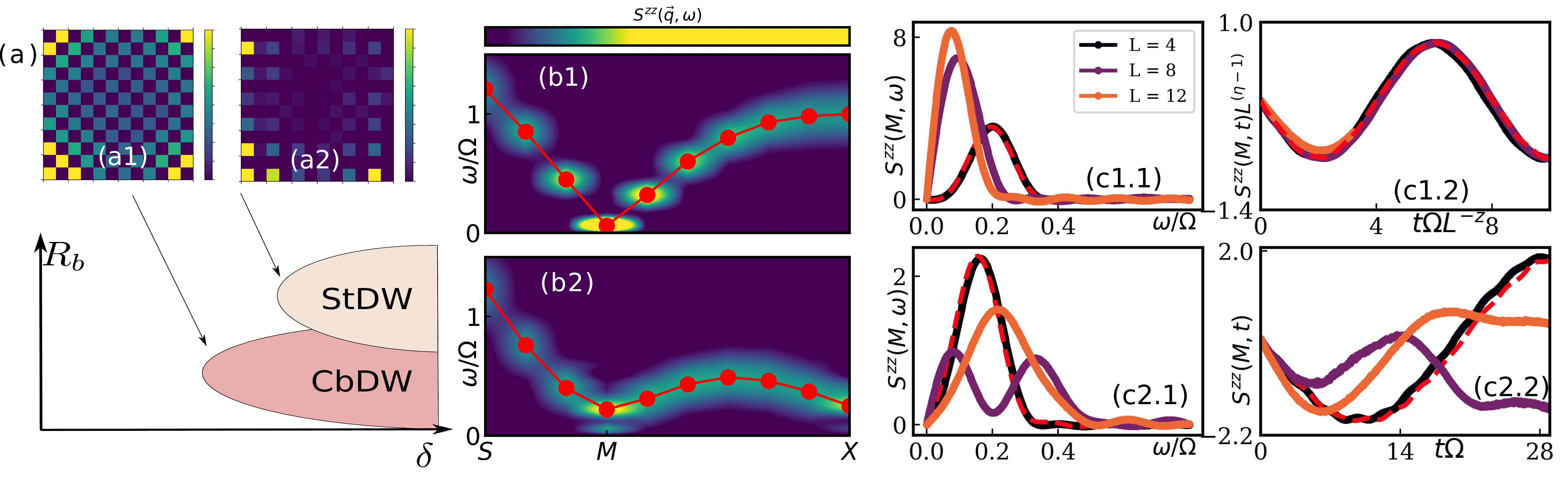}}}
\end{center}
\caption{\textit{Two-dimensional Rydberg atom arrays.} In panel (a), we show a schematic phase diagram of the RyM and the color plot of the spin correlation function,  $C^{z}(\vec{r}) = \left<\left< \sigma^z_{\vec{0}} \sigma^z_{\vec{r}} \right>\right>$, characteristic of CbDW and StDW phases; we consider (a1) $R_b=1.2$, $\delta = 1.05$ and (a2) $R_b=1.45$, $\delta = 2.00$.
In panels (b1) and (b2), we show the corresponding DSF for $L = 12$ along the following path of the Brillouin zone: $S = (\pi/2,\pi/2) \to M = (\pi,\pi) \to X = (\pi,0)$. 
The circle (red) points mark the peak of the DSF for each value of $\vec{q}$.
Finally, considering the same set of parameters $R_b$ and $\delta$, panels (c1.1,c1.2) and (c2.1,c2.2) show the respective finite-size scaling of the DSF and the dynamical correlation at the $\vec{q} = \vec{M}$ point; 
the dashed lines are exact results for $L = 4$.
}
\label{fig4} 
\end{figure*}

\textit{Results I: Two-dimensional Quantum Ising model.}
To benchmark our approach,  we consider the paradigmatic 2D QIM on a square lattice: 
 \begin{equation}
 \hat H = - \sum_{\left< i,j \right>} \hat{\sigma}^{z}_i  \hat{\sigma}^{z}_j - g \sum_{i} \hat{\sigma}^x_i.
 \label{QIM}
 \end{equation}
This model describes a second-order QCP at  $g_c  \approx 3.044$, separating a ferromagnetic from a paramagnetic phase. 
In the vicinity of $g_c$,  the major contributions to $S^{zz}(\vec{q}, \omega)$ 
come from low-energy quasiparticle excitations whose  frequency and spectral weight are expected to scale  as $\omega \sim L^{-z}$ and 
$\tilde{S} \sim L^{(1-\eta)}$, respectively \cite{Sachdev2011};
here, $z=1$ and $\eta \approx 0.04$ are critical exponents of the 3D-Ising universality class \cite{Pelissetto2002}.

We start discussing the DSF obtained with the real-space scheme shown in Fig. \ref{fig1}(c1). The plots show the DSF along a path through the first Brillouin zone for a lattice of size $N_s=10\times10$ at transverse field $g=3.1$. 
The main feature of $S^{zz}(\vec{q}, \omega)$ is a minimum at the $\Gamma$ point, corresponding to the low-energy gap.
We compare our results to series-expansion results up to fourth order in $g^{-1}$ \cite{Hammer2006,Hammer2006v2} and find very good agreement almost everywhere in the Brillouin zone. 
The most notable deviation appears at the gap closing point when approaching the critical value $g_c$, which we will discuss in more detail below.
Figure~\ref{fig1} (c2) shows similar results obtained with the momentum scheme in the vicinity of the $\Gamma$ point.
The fact that we can in this case use more efficient NQS architectures with built-in translational invariance allows us to simulate substantially larger system sizes up to $N_s=20\times20$ lattice sites and we again find good agreement with the series expansion.

Let us now focus on the behavior of the DSF in the vicinity of the QCP.
As a demanding benchmark for the accuracy of the NQS approach, we investigate the finite-size scaling of the spectral gap extracted from the DSF, $\Delta(\vec{q}=\Gamma)=\text{argmax}_\omega S^{zz}(\vec q=\Gamma,\omega)$. The expected universal scaling behavior $\Delta(\vec{q}=\Gamma) \sim L^{-z}$, with $L$ the linear system size, can be compellingly confirmed by our finite-size simulations up to $N_s=24\times24$; we obtain $z = 1.00 \pm 0.05$ by fitting such universal scaling, see SM \cite{SM}. We emphasize that our results show that with our NQS approach such large system sizes required to extract this universal behavior have now become within reach, while simultaneously also the diverging time scale associated with the closing of the spectral gap can be captured.

In Fig.~\ref{fig3}(c) we moreover show the frequency-resolved DSF at a few momentum points at $g=2.9$, i.e. on the ferromagnetic side of the QCP. These cuts reveal a double peak structure of $S^{zz}(\vec{q}, \omega)$, in particular an excitation at very low energy for $\vec q=\Gamma$. This feature is a signature of the $\mathbb Z_2$ symmetry-broken phase, where the degeneracy of the ground state is lifted to an exponentially small energy gap due to the finite system size.

\textit{Results II: Rydberg atom arrays.}
Now, to take our approach to a next level, beyond the QIM benchmark, we consider a long-range interacting model describing Rydberg atoms arrays on a square lattice (RyM).
The RyM  is defined as
\begin{equation}
 \hat H = - \frac{\Omega}{2} \sum_{i} \hat \sigma^x_i - \delta \sum_{i} \hat n_i  + \sum_{ i,j } \left(\frac{R_b}{r_{i,j}}\right)^6 \hat n_i \hat n_j,
\end{equation}
where $\hat n_i \equiv (\hat \sigma^z_i + 1)/2$. The parameter $\Omega$ represents the Rabi frequency, while $\delta$ denotes the detuning.
The term $ \left(R_b/r_{i,j}\right)^6 $ represents the interaction between atoms in Rydberg states,
$r_{i,j}$ being the distance between atoms at sites $i$ and $j$, and  $R_b$ the so-called Rydberg blockade radius \footnote{Here, we consider a finite cutoff for the interaction potential such that the interactions are set to $0$ for $r_{i,j} > r_0 = 3a$.}.

Recently, quantum simulation experiments \cite{Ebadi2021,Scholl2021} have motivated theoretical efforts to understand ground-state properties of the square-lattice RyM \cite{Samajdar2020, Felser2021, Kalinowski2022}. The interplay between the parameters $\delta$ and $R_b$ leads to density-wave phases and related QCPs. As an example, we present a schematic phase diagram for RyM in Fig. \ref{fig4}(a), indicating the emergence of checkerboard (CbDW) and striated (StDW) density-wave. Ground state searches and the estimations for the disordered-CbDW and disordered-StDW transitions are discussed in the SM \cite{SM}, which is consistent with the literature \cite{Kalinowski2022}.
Here, we focus in computing spectral functions in the vicinity of QCPs.

Near the disordered-CbDW QCP, $S^{zz}(\vec{q},\omega)$ is characterized by a dominant low energy mode at $\vec{q} = \vec{M}$, see Fig. \ref{fig4} (b1). The finite-size scaling of  the DSF (we obtain $ z = 0.9 \pm 0.2$ by fitting the size scaling of the the lowest-energy gap, see SM \cite{SM}) and the dynamical correlator $S^{zz}(\vec{M},t)$ is consistent with a second-order QCP in the same universality class as the previously discussed 2D QIM; see Fig. \ref{fig4} (c1.1) and (c1.2). 
 
Close to the disordered-StDW QCP, however, $S^{zz}(\vec{q},\omega)$ exhibits qualitatively different behavior. It is characterized by two dominant low energy modes, occurring at $\vec{M}$ and $\vec{X}$; see Fig. \ref{fig4} (b2). In addition, the spectral weight associated with the lower-energy peak of $S^{zz}(\vec{q}=\vec{M},\omega)$ decreases with system size, while we observe a spectral weight transfer to higher energies for larger values of $L$; see Fig. \ref{fig4} (c2.1). These results are consistent with the prediction of a first-order QCP \cite{Kalinowski2022}. 
We note, however, that for $L = 12$ our simulations cannot resolve a lower energy peak. 
For first-order phase transitions, the lowest energy gap is expected to vanish exponentially as $L$ increases \cite{Vicari2014}. This implies that to resolve the spectral gap accurately, we have to perform simulations up to an exponentially long time, $t_{max}$, and with a Fourier transform done with a Gaussian broadening factor scaling as $1/t_{max}$ \cite{SM}. In this regime, the phase transition is better evidenced by ground-state static properties related to the DSF \cite{Kalinowski2022}.



\textit{Discussion and conclusions.}
In summary, we have proposed a cutting-edge method to simulate spectral properties of 2D quantum many-body systems, which relies on the representational power of NQSs to access  spectral functions during the real-time dynamics of local excitations. We demonstrated that this scheme allows us to reliably perform finite-size scaling of dynamical properties near 2D quantum critical points for unprecedented system sizes. 
A promising future direction is to characterize spectral features of 2D models stabilizing quantum spin liquid phases.
Of particular current interest are spin liquids with Ising-like interactions, such as the 2D RyM in a ruby lattice, which has recently been proposed as a way to realize $\mathbb Z_2$ topological order states in programmable quantum simulators \cite{Semeghini2021,Ruben2021}, and also Kitaev-type spin models \cite{Kitaev2006}, where highly-accurate simulations of DSF are essential to characterize exotic fractionalized excitations \cite{Wessel2018,Meng2018}.

\begin{acknowledgments}
 
\paragraph{Acknowledgements. -} We thank Matteo Rizzi, Michael Knapp and Ao Chen for fruitful discussions. 
This project has received funding from the European Research Council (ERC) under the European Union’s Horizon 2020 research and innovation programme (grant agreement No. 853443).
MS was supported through the Helmholtz Initiative and Networking Fund.
The authors gratefully acknowledge the Gauss Centre for Supercomputing e.V. (www.gauss-centre.eu) for funding this project by providing computing time through the John von Neumann Institute for Computing (NIC) on the GCS Supercomputer JUWELS \cite{JUWELS} at J\"ulich Supercomputing Centre (JSC).
We used the jVMC codebase \cite{Schmitt2022} that is built on the JAX library \cite{jax2018github} to implement our approach. We use the quspin package \cite{Bukov2017,Bukov2019} to obtain the exact diagonalization results.
The data shown in the figures are available on Zenodo \cite{mendessantosData}.

\end{acknowledgments}


\phantomsection
\addcontentsline{toc}{chapter}{Bibliography} 
\bibliography{Refs}

\newpage

\section{Further details on the simulation of dynamical correlations with tVMC}
In this section, we discuss further important details  to simulate the dynamical correlations  $S^{zz}(\vec{q},t)$ and $G^{zz}_{\vec{r},\vec{r}^{\prime}}(t)$.

\textit{Momentum-space scheme.} As described in the main text, we obtain $S^{zz}(\vec{q},t)$ by (i) performing a two-sided time evolution followed by (ii) the evaluation of the overlap of two  time-evolved states; see Eq. \eqref{eqOverlap}.

First, to time evolve the states  $\bra{\phi^z_{\vec{q}}(-\delta t)}$ and $\ket{\phi^z_{\vec{q}}(\delta t)}$, we solve the TDVP equation [Eq. \eqref{eq:tdvp}]  with a second-order integration method. Specifically, we implement the adaptative-time-step scheme to describe the time evolution of the ket-state $\ket{\phi^z_{\vec{q}}(\delta t)}$. 
Such an approach allows us to estimate errors based on varying time step sizes 
and adjust  $\delta t$ during the simulation. The adaptative-time-step scheme is implemented with the Heun method; we refer the reader to ref. \cite{Schmitt2020} for further details.
The time evolution of the bra-state $\bra{\phi^z_{\vec{q}}(-\delta t)}$ is then performed by using the Heun method with the time step $\delta t$ defined by the ket-state time evolution.

Second, to compute the overlap between the two unnormalized states  $\bra{\phi^z_{\vec{q}}(-t/2)}$ and $\ket{\phi^z_{\vec{q}}(t/2)}$  we employ  Monte Carlo sampling to  obtain the normalized overlap (to simplify the notation, let us define $\bra{\phi^z_{\vec{q}}(-t/2)} = \bra{\phi_1} $ and $\ket{\phi^z_{\vec{q}}(t/2)} = \ket{\phi_2} $)
\begin{equation}
 O = \frac{\braket{\phi_1 | \phi_2}}{\sqrt{\braket{\phi_1 | \phi_1}\braket{\phi_2 | \phi_2}}}.
\end{equation}
The idea is that the quantity $p_1 = \braket{\phi_1 | \phi_2}/\braket{\phi_1 | \phi_1} $, can be estimated via Monte Carlo sampling, i.e.,
\begin{equation}
 p_1 = \sum_{\vec{s}}  \frac{|\phi_1(\vec{s})|^2}{\braket{\phi_1 | \phi_1}}  \frac{\phi_2(\vec{s})}{\phi_1(\vec{s})} \approx \frac{1}{N_{MC}}\sum_{i} \frac{\phi_2(\vec{s}_i)}{\phi_1(\vec{s}_i)},
\end{equation}
where $\vec{s}_1, ..., \vec{s}_{N_{MC}}$ represent a set of $N_{MC}$ MC samples generated by the Born distribution $|\phi_1(\vec{s})|^2$.
Analogously,   $p_2 = \braket{\phi_2 | \phi_1}/\braket{\phi_2 | \phi_2} $ is estimated via MC sampling of  $|\phi_2(\vec{s})|^2$.
By combining the two results, we obtain $O = \sqrt{p_1 p_2^*}$ \cite{Wu2020}.

Finally, by computing $O_{\vec{q}} = \braket{\phi^z_{\vec{q}}(-t/2) | \phi^z_{\vec{q}}(t/2)}/N_{\vec{q}}$,
where the normalization constant 
\begin{equation}
N_{\vec{q}} = \sqrt{\braket{\phi^z_{\vec{q}}(-t/2) | \phi^z_{\vec{q}}(-t/2)} \braket{\phi^z_{\vec{q}}(t/2) | \phi^z_{\vec{q}}(t/2)}} 
\end{equation}
is simply related to the ground-state static structure fractor, $N_{\vec{q}} = N_s \sum_{\vec{r}} e^{i \vec{r} \vec{q}} \left< S^{z}_{\vec{r}} S^{z}_{\vec{0}} \right> $, we obtain the dynamical correlation in momentum space 
\begin{equation}
 S^{zz}(\vec{q},t) = e^{iE_0t} N_{\vec{q}} O_{\vec{q}}.
\end{equation}

\

\textit{Real-space scheme.} We use Eq. \eqref{localob} to compute the longitudinal component of the Green function $G^{zz}_{\vec{r},\vec{r}^{\prime}}(t)$,
for a deduction of such expression see refs. \cite{Knap2013,Baez2020}.
The expectation value  $\bra{\phi_{\vec r'}^z(t)}\hat S_{\vec r}^z \ket{\phi_{\vec r'}^z(t)}$ is obtained by simulating the 
time evolution of the local excitation $\ket{\phi_{\vec r'}^z(t)} = e^{-i \hat H t} e^{i\frac{\pi}{4} \hat{S}_{\vec{r^{\prime}}}^z} \ket{\psi_0}$ with the tVMC, where
the local excitation is encoded on the NQS ansatz by adding the configuration-dependent factor $f(\vec{r^{\prime}},\vec{s}) = 1/\sqrt{2} \left( 1 - i s_{\vec{r^{\prime}}} \right)$ on the initial-time quantum state
\begin{equation}
 \ket{\phi_{\vec r^{\prime}}^z(0)} = \sum_{\vec{s}}  f(\vec{r^{\prime}},\vec{s}) \psi_{\vec{\theta}}(\vec{s}) \ket{\vec{s}}.
\end{equation}
Similarly, the term $\bra{\psi_0}\hat S_{\vec r}^z\hat S_{\vec r'}^z(t)\hat S_{\vec r}^z \ket{\psi_0}$ is obtained by simulating the time evolution of
the local excitation $ \hat{S}_{\vec r}^z \ket{\psi_0}$, which is encoded on the NQS ansatz by adding the factor $f^{\prime}(\vec{r},\vec{s}) =  s_{\vec{r}}$.

\

\textit{Fourier transform}
Let us now discuss the details of the Fourier transform (FT) used to obtain the dynamical structure factor in the frequency-momentum domain: $S^{zz}(\vec{q},\omega)$.

In the momentum-space scheme,  we perform a  FT of $S^{zz}(\vec{q},t)$ to the frequency domain
\begin{equation}
 S^{zz}(\vec{q},\omega) = \frac{1}{2\pi} \int_{-\infty}^{\infty} dt e^{i\omega t} e^{- \gamma^2 t^2}  S^{zz}(\vec{q},t),
 \label{FT}
\end{equation}
where we apply a Gaussian envelope to avoid the finite-time effects of our simulation;
the parameter $\gamma = 2/t_{max}$ controls the  broadening factor of the Gaussian peaks,  with $t_{max}$  being  the maximum times reliably obtained during our simulations.
For the results shown in the main text, we typically performed simulations with $t_{max} \approx 20J$ and $t_{max} \approx 40\Omega$ for the 2D QIM and RyM, respectively. %
Furthermore, we employ the trapezoidal rule to perform the numerical integration of Eq. \eqref{FT}.

In the real-space scheme, we calculate the Green function, $G^{z}_{\vec r,\vec r'}(t)$, for all pairs of spins ($\vec r,\vec r'$), and then perform a FT in real space and time  to obtain $G^{z}(\vec{q},\omega)$ (the FT to the frequency domain is performed as described above). 
Finally, we relate the retarded Green function to the DSF via the fluctuation-dissipation theorem
\begin{equation}
 S^{zz}(\vec{q},\omega) = -\frac{1}{\pi} Im[G^{z}(\vec{q},\omega)].
\end{equation}

\section{Network architectures and hyperparameters}

As mentioned in the main text, the real-space-scheme is implemented with restricted Boltzmann machine (RBM) NQSs \cite{Carleo2017},
which are architectures composed by a single fully-connected (dense) hidden layer.
Concretely, RBMs  are defined as
\begin{equation}
 \psi_{\vec{\theta}}(\vec{s}) = \prod_{i=1}^{M} \cosh\left( \sum_{j=1}^N W_{i,j} s_j + b_i\right),
\end{equation}
where the set of variational parameters $\vec{\theta} = (W,\vec{b})$ are the complex weight matrix, $W$, and the hidden bias, $\vec{b}$.
The hyperparameter $M$ represents the number of nodes in the  single hidden layer, which sets the
total number of variational parameters, i.e., 
$N_p = 2(N M  + M)$.

For the results shown in Fig. \ref{fig1} (b1) and (c1) we consider $M = 40$.\\

The momentum-space scheme is implemented with the convolutional neural network ansatz \cite{Schmitt2020}.  
Specifically, we consider architectures composed of a set of $L_l$ stacked convolutional layers.
Each output neuron of a convolutional layer is described by
\begin{equation}
 a^{(l)}_{c,k}(s) = f_l \left( \sum_{c^\prime=1}^{\alpha_{(l-1)}}  \sum_{t=1}^{F^{(l)}} \tilde W^{(c,l)}_{k,c^\prime,t}  a^{(l-1)}_{c^{\prime},T_k(t)} + b_c^{(l)}   \right),
\end{equation}
where $a^{(l)}_{c,k}$ represent a value at a channel $c$ of an output layer $l$; the index $c = 1, ..., \alpha_l$ runs over the  total number of output channels.
The value of a neuron, $a^{(l)}_{c,k}$, is defined by an affine map that resembles a convolution of the previous layer with
a filter along an orbit generated by translations.
Specifically, the individual neurons $a^{(l)}_{c,k}$  are coupled with identical weights $\tilde W$ to the neurons of the previous layer,
which are transformed by translations, $T_k(t)$, by some fixed \textit{stride}.
In particular, we consider a $2D$ filter with size $F \times F$.
Such map can also include a bias $b_m^{(l)}$ for each channel. 
Finally, $f_l(...)$ represents an activation function.

Starting with the initial layer, which is the spin configuration $a^{l=0}_{1,j} (s) = s_j$, and considering the recursive relation for the $L_l$ convolutional layers,
we can define the logarithm of the wave function in terms of the last convolutional layer 
\begin{equation}
 \ln \psi_{\vec{\theta}} = \sum_{c,k} a^{(L_l)}_{c,k}(s).
\end{equation}

The total number of variational parameters a CNN NQS is given by
\begin{equation}
 N_p^{CNN} = 2 \left( F^2 \sum_{l=1}^{L_l} \alpha_{l-1} \alpha_l + N_b \right),
\end{equation}
where $\alpha_0 = 1$ and $N_b$ is the number of biases.
In this work, we consider CNN NQSs with the following structure $\vec{\alpha} = (\alpha_1,..., \alpha_{L_l};F)$, where
$\alpha_k$ is the number of channels in each layer $k$.
Furthermore, we use a stride $\vec s = [1,1]$.
For the results present in the main text we adopt the following hyperparameters:

\begin{figure}[t]
\begin{center}
{\centering\resizebox*{8cm}{!}{\includegraphics*{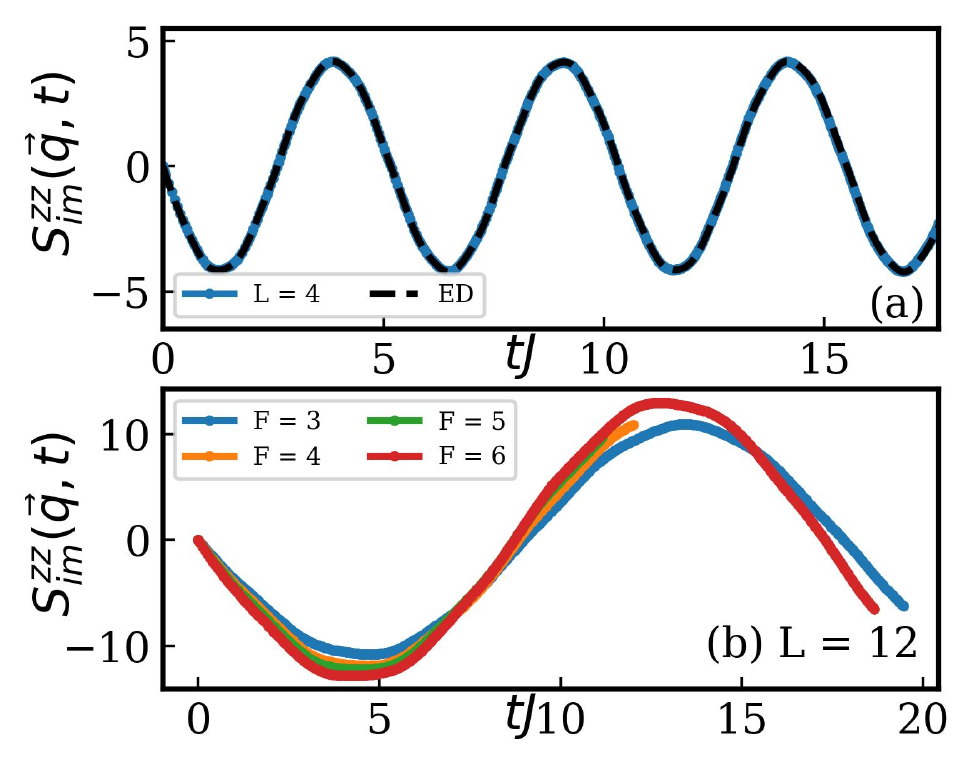}}}
\end{center}
\caption{\text{Additional results for the 2D QIM.} In panel (a) we show a comparison of the imaginary part of dynamical correlation $S^{zz}_{im}(\vec{q},t)$  with exat results for $g = 3.1$,
while in panel (b) we check the convergence of the results for  $L = 12$ by increasing the linear size of CNN filter $F$; in this case we consiter $\vec{\alpha}=(4,3;F)$ and $g = 3.044$.}
\label{fig1SM} 
\end{figure}

\begin{figure*}[t]
\begin{center}
{\centering\resizebox*{18cm}{!}{\includegraphics*{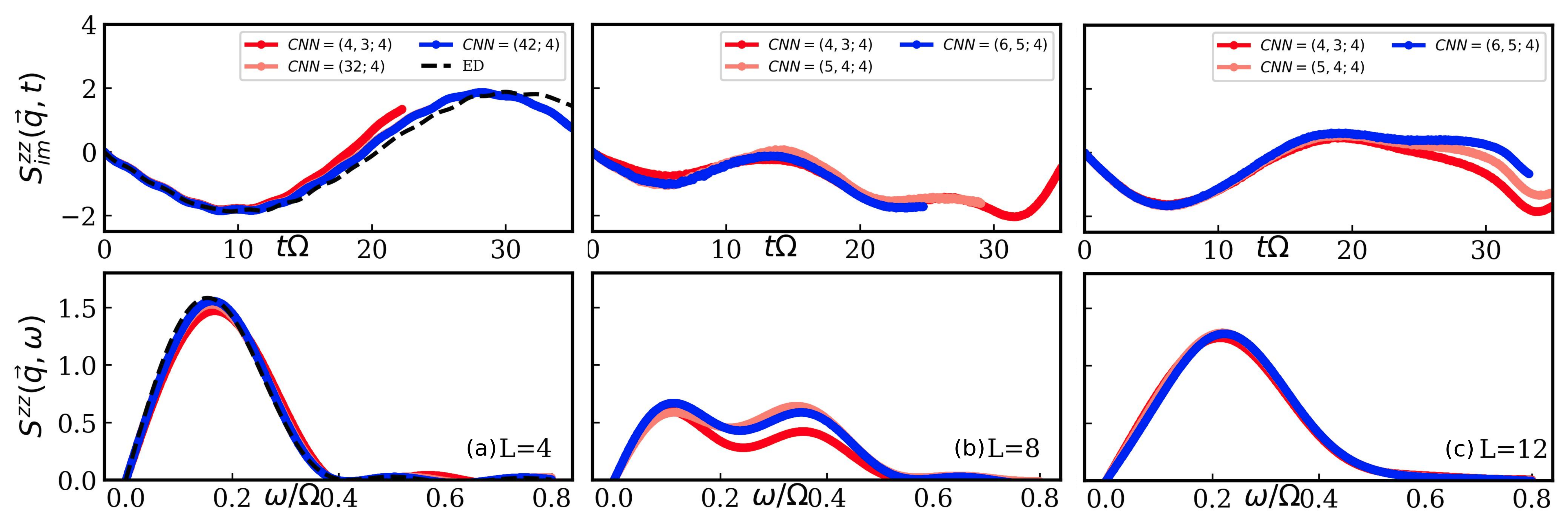}}}
\end{center}
\caption{\text{Additional results for the 2D RyM.} The upper pannel show $S^{zz}_{im}(\vec{q} = M,t)$ for (a) $L = 4$, (b) $L =8$ and (c) $L = 12$ and
$R_b = 1.45$ and $\delta = -2.00$. In the lower panels we show the correspond Fourier transform to frequency $S^{zz}_{im}(\vec{q} = M,\omega)$. }
\label{fig2SM} 
\end{figure*}

\begin{itemize}
 \item Results for the  2D QIM (Figs. \ref{fig1} and \ref{fig3}).
For $L < 20$, we consider $\vec{\alpha} = (4,3;F=L/2)$, while for $L = 20$, $\vec{\alpha} = (4,3;F=8)$, and $L = 24$, $\vec{\alpha} = (4,3;F=10)$.
We employ a sixth-degree polynomial in the first layer and a fifth-degree polynomial in subsequent layers.

\item Results for the  2D RyM (Fig. \ref{fig4}). We consider: $\vec{\alpha} = (6,5;F=4)$ for $L=12$ and $\vec{\alpha} = (4,3;F=4)$ for $L = 8$ and $4$ results. We use fifth-degree polynomial activation functions in all the CNN layers. 

\end{itemize}

We perform all the simulations  considering $N_{MC} = 4 \times 10^4$ Monte Carlos samples.

\section{Convergence checks and bechmarks}

We now provide additional results to check the convergence of the results presented in Figs. \ref{fig3} and \ref{fig4} of the main text. Furthermore, we perform   further benchmarks  with exact results and theretical descriptions of QCPs. 

For the smaller system size considered here, $L = 4$, we make a direct comparison with exact results. 
As can be seen from Fig. \ref{fig1SM} (a), $S^{zz}_{im}(\vec{q} = \Gamma,t)$ exhibit an almost perfect agreement with ED results.
To be more quantitative in the comparison,  we present the values of the lowest energy gap, $\Delta(\vec{q}=\Gamma) $ computed with NQS and ED on table \ref{table1}. 
In the exact case, we directly compute the gap from $\Delta(\vec{q}=\Gamma) = E_1 - E_0$, where $E_0$ and $E_1$ are the ground-state and the first-excited-state energies, respectively. Using the NQS approach, we extract the gap from the DSF, i.e., $\Delta(\vec{q}=\Gamma)=\text{argmax}_\omega S^{zz}(\vec q=\Gamma,\omega)$.
The discrepancy between such results are $\epsilon = |\Delta - \Delta^{EX}|/ \Delta^{EX}  < 0.2\%$ for all the values  $g$ considered in Fig. \ref{fig3}, which is within
the statistical error bars for $N_{MC} = 4 \times 10^4$ samples used here.

For $L > 4$, a direct comparison with exact results is not possible. 
Nevertheless, an important aspect of the neural quantum state approach is that we can test the accuracy of the results by comparing NQS with different sizes or architectures.
For this,   we provide additional results for  $S^{zz}_{im}(\vec{q} = \Gamma,t)$ and different $\vec{\alpha} = (4,3;F)$. By focusing on the  challenging regime $g = 3.044$,
we observe the convergence of the results for the larger values of $F$ considered; see Fig. \ref{fig1SM} (b).

\

\begin{table}
\begin{tabular}{ |p{1.5cm}||p{2cm}|p{2cm}|p{1.5cm}|  }
 \hline
 \multicolumn{4}{|c|}{Spectral gap (QIM) - L = 4} \\
 \hline
   --  &   Exact & NQS   & $\epsilon$ \\
 \hline
 $g = 3.5$   & $2.1124$   &  $2.1138$  &   $0.0007$\\
 \hline
 $g = 3.1$   &  $1.2159$  & $1.2178$   & $0.0016$\\
 \hline
 $g = 3.044$   &   $1.0968$ & $1.0970$   & $0.0001$\\
 \hline
\end{tabular}
\caption{Spectral gap of the 2D QIM.}
\label{table1}
\end{table}
\

We now consider the convergence of our results for the Rydberg-atoms-array model (RyM).
In particular, we focus on a regime of parameters in which the system is near the disordered-StDW transition (i.e., $R_b = 1.45$ and $\delta = -2.00$). In this case,  we observe that $S^{zz}_{im}(\vec{q} = M ,\omega)$ exhibit a non-trivial size scaling, as discussed in Fig. \ref{fig4} (c2) of the main text
the  spectral weight associated with the lower-energy peak of $S^{zz}(\vec{q}=\vec{M},\omega)$ decreases with system size,
while we observe a spectral weight transfer to higher energies for larger values of $L$; see Fig. \ref{fig4} (c2.1). 
As can be seen in Fig. \ref{fig2SM}, results for $S^{zz}_{im}(\vec{q} = M,t)$ for different $\vec{\alpha}$ converge for $t \lesssim 30\Omega$. 
By chosing $t_{max} = 30\Omega$, we show in the lower panels of \ref{fig2SM} the results in frequency, confirming the non-trivial scaling of $S^{zz}_{im}(\vec{q} = M ,\omega)$.

\begin{figure}[t]
\begin{center}
{\centering\resizebox*{8.5cm}{!}{\includegraphics*{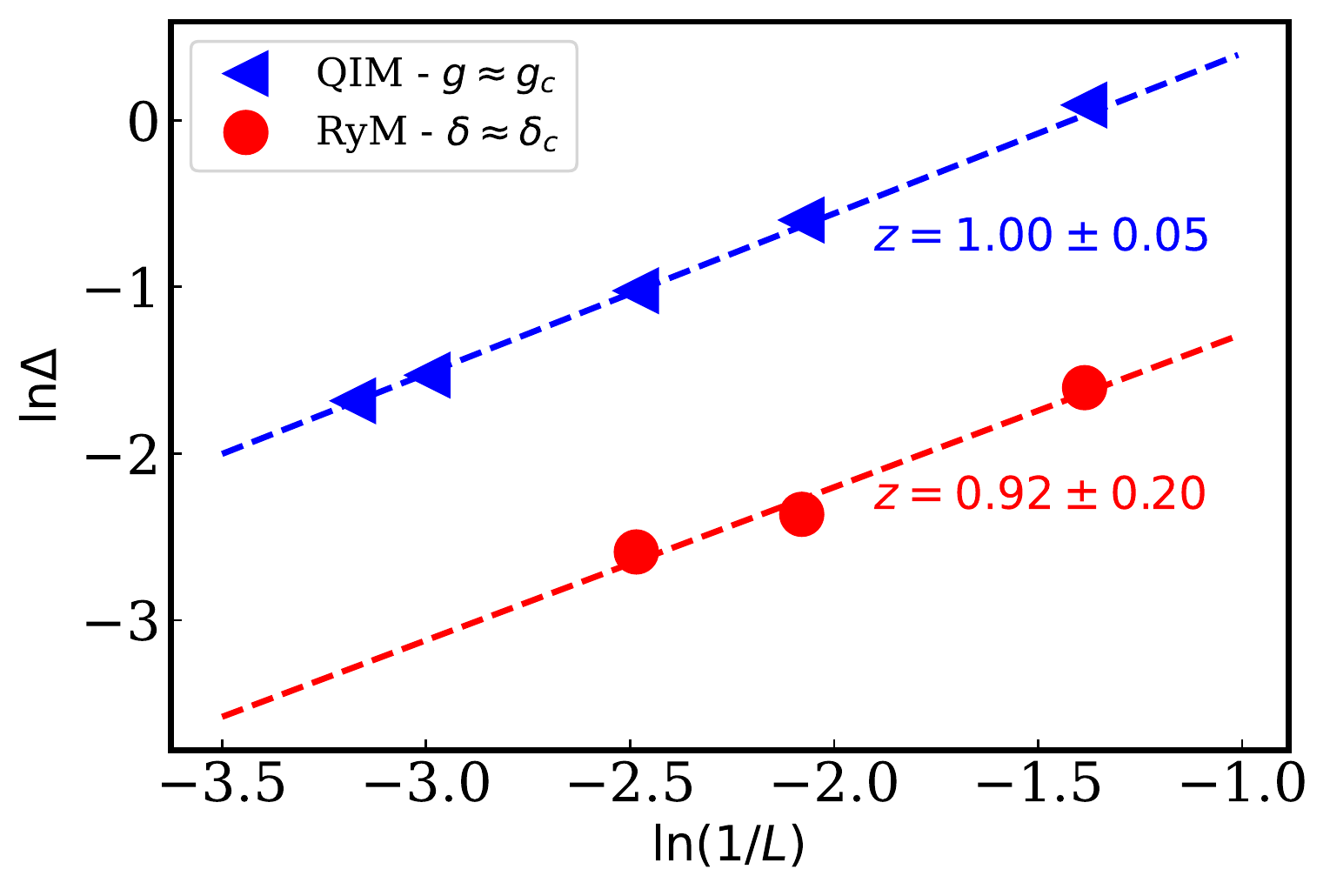}}}
\end{center}
\caption{Data for the finite-size dependence of the spectral gap $\Delta$ as obtained through the NQS simulations close to both the QIM and the disordered-CbDW quantum critical points. We further include the fitted values of the critical exponent $z$ obtained by assuming that $\Delta \sim L^{-z}$ and the standard errors of the coefficient of the linear fitting. }
\label{fig4SM} 
\end{figure}

Finally, as a more demanding benchmark of our results, we perform a quantitative verification of the size-scaling of the low-energy spectral gap, $\Delta \sim L^{-z}$, at the QCPs considered in this work. 
Fig. \ref{fig4SM} displays our data for $\Delta$ as a function of system size. Further, we also show fitted values of $z$ for both the QIM (for $g = 3.044$) and the disordered-CbDW (for $R_b = 1.2$ and $\delta = 1.05$) transitions obtained by assuming a functional dependence $\Delta \sim L^{-z}$ of the gap. Our estimation of  $z$ is consistent with the theoretical expectations for the 2D QIM, $z = 1$. Going beyond a benchmark, we also estimate the dynamical exponent for the disordered-CbDW [in this case, we consider $\Delta(\vec{q}= \vec{M})$]  the obtained value is consistent with the Ising universality class \cite{Samajdar2020,Kalinowski2022}.

\section{Ground state results for the 2D RyM}

\begin{figure}[t]
\begin{center}
{\centering\resizebox*{8.5cm}{!}{\includegraphics*{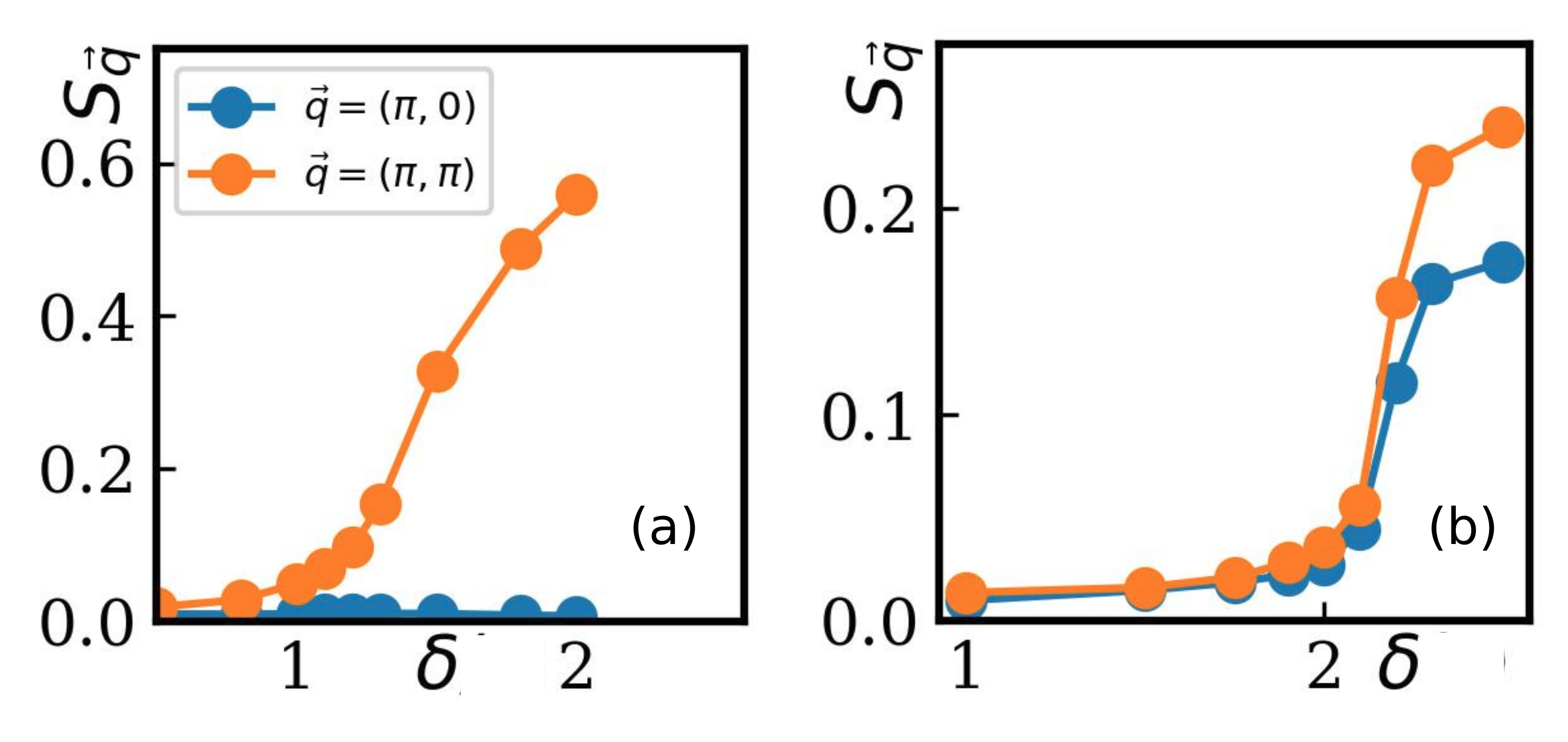}}}
\end{center}
\caption{\text{Ground-state results for the 2D RyM.} Static structure factor $S_{\vec{q}}$ as function of $\delta$ for (a) $R_b = 1.2$ and (b) $R_b = 1.45$, and $L = 8$. }
\label{fig3SM} 
\end{figure}

 As a complement to the results of the 2D RyM,
we discuss the onset of the CbDW and StDW phases by 
considering the ground-state (GS) correlations
\begin{equation}
C^{z}(\vec{r}) = \left<\sigma^z_{\vec{0}} \sigma^z_{\vec{r}} \right> - 
\left<\sigma^z_{\vec{0}} \right>\left< \sigma^z_{\vec{r}} \right>.
\end{equation}
Particularly, we show in Figs. \ref{fig3SM} (a) and (b)  the behavior of
the static structure factor, $S_{\vec{q}} = 1/N_s \sum_{\vec{r}} e^{i\vec{r} \vec{q}}C^{z}(\vec{r}) $, along the lines $R_b = 1.2$ and $R_b = 1.45$, respectively. 
For $R_b = 1.2$, we note an increase of the $S_{\vec{q = M}}$ for $\delta \approx 1.05$, indicating the emergence of a phase with CbDW pattern of correlations, while for $R_b = 1.45$ we note an increase of  both
$S_{\vec{q = M}}$ as $S_{\vec{q = X}}$ for $\delta \approx 2.1$, indicating the emergence of the StDW phase. Those results are in agreement with previous simulations of GS properties the 2D RyM \cite{Samajdar2020, Kalinowski2022}.

\end{document}